# On Population Kinetics of an Aging Society:
## Aging and Scurvy

John T.A. Ely, PhD*,
Radiation Studies, Box 351310,
University of Washington , Seattle, WA  98195

**Summary.**  Evidence from skeletal remains and pictorial records for the past several millennia established  that humans have had a very inadequate ascorbic acid intake and commonly exhibited signs of frank scurvy, as well as the invisible signs of subclinical scurvy (absence of AA in scorbutic urine is universal).  They suffered scurvy's mortality, being a major cause of death especially in times of stress due to dietary fluctuations, climate, migrations, voyages, battles, etc.  Rapid aging and death by acute scurvy, and infectious and degenerative diseases related to chronic scurvy added to the toll.

The loss of function and changes in appearance that we see in humans in the second 50 years of life are all essentially simple to avoid. Evidence is very strong that people whose diet contains all the essential nutrients in amounts which have been shown to optimize life can reduce signs of aging in the second 5 decades of life.  However, this is very rare because there is almost no one who does not eat the refined diet of the affluent societies.  This diet contains so much white sugar, white bread, and other refined carbohydrate that blood sugar is always elevated.  When blood sugar is high, the sugar monopolizes the insulin and prevents insulin mediated active transport of AA into cells;  buffy coat AA is low, immunity and all metabolic processes, including maintenance and repair of structural proteins, are impaired.  The bottom line is that the loss of function and deterioration in appearance of aging can be greatly slowed (see Ely JTA, Biochemical Reversal of Aging. APS Baltimore, March 2006; http://meetings.aps.org/Meeting/MAR06/Event/42176). These two properties of aging are common signs of scurvy.  Therefore, one is compelled to accept the statement that "aging is scurvy"!

We present here an overview of a vast array of evidence on the connections between aging and scurvy  (1, Ch 8,9). Scurvy results from an inadequate intake of ascorbic acid (AA),  frequently called Vitamin C although it is not a vitamin.  Vitamins are nutrients needed in only minute amounts. But, for optimum health, the average unstressed human requires  over 4 grams of ascorbic acid daily. Scurvy has strongly affected western humans since the change from a Paleolithic fare, providing ~400 mg AA daily, to a woefully inadequate agrarian diet with only ~80 mg (2).  Neither of these amounts would be adequate AA for the maintenance of youthful structural proteins "unquestionably" necessary for slow aging and long life (3,4).  Both diets would produce scorbutic states in humans. The first state would be a chronic biochemical impairment with minimal visible signs. In the second state people would appear much older and be significantly more prone to infectious and degenerative diseases, rapid aging and life spans shortened by scurvy as principal cause of death. These effects are so logically compelling they make an even stronger case for the statement "aging is scurvy".

---

*Corresponding author:  John T.A. Ely, current address, PO Box 1925, Palmerston North, New Zealand. ely@u.washington.edu

**Doubters and Adversaries.** There is a widely held and fallacious opposing view, that "AA is a vitamin". This view is easily identified as a principal cause of early aging and death in affluent western countries such as the US, especially for those on refined (high sugar) diets (5). The very strong effect of the high glycemic level of affluent societies in antagonizing AA utilization has been summarized (3, 6). According to Pauling, US so-called "authorities" have asserted that AA is a vitamin (1, p. 99; we urge that everyone read pp 394-8 on Pauling's vast but little-known scientific accomplishments): "these authorities persist in ignoring the many studies that have been carried out demonstrating that an intake of several grams per day leads to improved health."

**Postponing and or Reversing Aging.** In humans, signs of aging include crosslinking, loss of protein solubility, glycation, reduced skeletal mass, frailty, stooped posture, etc. These all become prominent between ages 55 and 85 when about half of all deaths after "maturity" occur; for "maturity", we use age of first estrus (i.e., 11 years). In humans, only a few percent live to age 90 (7). The ratio of human life expectancy (~88 years) to age of maturity is only 8. Among many normal (AA synthesizing) mammals, this ratio is much larger (i.e., ~25). These mammals have no gross changes in appearance and never have scurvy until the end of life when their AA synthesis declines rapidly (8). Precipitous aging and death can be postponed by exogenous AA (a possible practice in zoos with high value mammals). The simplest way to slow aging in humans is by carefully correcting the diet to include the optimal intakes of AA and all other essential nutrients, especially those such as lysine (caution re excess) that are necessary for replacement of the structural proteins collagen and elastin and for opposing other age-related changes. To reverse aging, AA intake should be increased to the maximum level not exceeding bowel tolerance. The other competitors for insulin must be reduced to those necessary for metabolic function. An example of aging reversal by our biochemical method can be seen on the website (http://faculty.washington.edu/ely/arxivaging.html).

1. <u>Scurvy, AA, Structural Proteins and Aging.</u> For over 4 decades, marked differences between connective tissue of young and old have been studied in humans and other mammals (4, 9). With increasing age there is a rising rate of structural protein degeneration. This is a principal sign of aging. Also, AA levels decrease (ie, scurvy increases) with increasing age (10). Thus, the body's ability to synthesize, repair and replace structural proteins to prevent the loss of youthful flexibility and elasticity that is incurred by their deterioration is diminished. Collagen becomes tougher, more crystalline and less soluble. Elastin increases calcification, losing elasticity and fragments with age. Ground substance deteriorates by increasing density and aggregation. To maintain youthful characteristics of structural proteins requires optimal intakes of AA and the other essential nutrients, especially lysine. High levels of AA are needed to support the hydroxylation requirements in conversion of proline and lysine for collagen and elastin. Elastin is particularly rich in lysine, four molecules of which combine to form desmosine, the specialized linkage that enables the elastic net-like properties of elastin. Loss of the youthful properties of structural proteins results in arterial rigidity, increased systolic pressure, reduced cardiac ejection fraction, and the various heart diseases known to increase greatly in aging.

In 1991, Linus Pauling reported (11) the striking recovery of a hopeless no longer operable terminal heart disease patient, the first to adopt Pauling's newly proposed concept of high lysine and AA (with other essential nutrients). For convenience of the reader, with permission of the editor, we have displayed this article on a website (11). In March 1991, a 71 year old retired biochemist patient who had continuously worsened for 33 years in spite of 3 bypasses (and much other surgery) was informed most of the vessels were clogged and he could not have additional surgery. In May, with steadily increasing pain and deterioration, he elected to start Pauling's lysine/AA dietary therapy at a low dose. By mid June, he was taking the full therapeutic dose of 5 to 6 grams lysine and other essential nutrients. By July, he was free of angina, could even walk two miles and do yard work without pain. In August, he cut up a tree with a chain saw, and then started painting his house. By October, after some adjustments in regimen, he was free of all disease symptoms and was able to reduce nutrient dosing to a maintenance level (which, it has been suggested, all such patients should continue for life). In spite of age >70, after raising lysine to cure his heart disease this patient was now capable of vigorous exertions typical of a youthful human because high AA intake restored youthful elasticity to his structural proteins (lungs, vasculature, etc).

2. <u>Scurvy, AA, Glycation and Aging.</u>
The concept of protein glycation, another most important aging mechanism, was developed by Cerami et al (12). There is a positive association between glycemic level and the rate of glycation. One of Ely's discoveries was that AA opposes glycation, and, as a result, there is a monotonic inverse association between ascorbic acid level and glycation rate at a given plasma glucose level (13-16). This antagonism of glycation is mediated by the ascorbate anion displacing glucose from its transient binding to the amine before the Amadori rearrangement makes the glycation irreversible. Thus, the glycation of all proteins, including the wrinkling of skin, is a sign of aging and proceeds at a rate that is higher when plasma AA levels are inadequate to prevent or oppose it.

3. <u>Scurvy, AA, Radiation and Aging</u>. Exposure to radiation accelerates aging and decreases life expectancy (4, 17). Many authors have long reported that AA protects against radiation (18), and that radiation decreases AA levels in vivo (19, 20). It was shown nearly 40 years ago that AA protects enzymes (such as lysozyme, aldolase, etc) against inactivation by ionizing radiation. Thus, high AA opposes the aging effects of radiation, but the low levels in scurvy accelerate aging and death.

4. <u>Scurvy, AA, Free Radicals and Aging.</u> Denham Harman's free radical theory (7, 21) is generally accepted as describing one of the most important mechanisms of aging. He suggested the similarity between certain aspects of radiation injury and aging may reflect common physiological and cellular injuries (21). Or, analogous chemical events such as free radical reactions may occur in both radiation induced and normal aging. These concepts have been helpful in explaining the biological effects of radiation (22). The essential nutrients, AA plus vitamin E, provide most important free radical protection without which aging is very rapid, as shown by the symptoms of scurvy. Evidence suggests, because autooxidation occurs in senescent tissues, scurvy must be avoided and maintenance of high AA levels is a necessary preventive therapy. According to Tappel (23), "Optimum amounts of ascorbic acid would be important in any attempts to slow the aging processes".

5. <u>Scurvy, AA, Immunity and Aging.</u> The hexose monophosphate shunt rate was discovered to be strongly proportional to intracellular AA concentration in 1971 (24). In support of cell mediated immune response, the hexose monophosphate shunt supplies ribose needed for mitosis in expanding a clone of lymphocytes, and hydrogen-peroxide required for phagocytic activity. If AA is low, as in humans and other AA non-synthesizers and in all aged mammals, the shunt rate and cell mediated immunity fall. Inadequate AA not only affects cell mediated immunity directly, but also indirectly as a result of the high glycemic levels characteristic of affluence; glucose competitively inhibits the insulin-mediated active transport of AA into cells resulting in a clearly depressed buffy coat AA level (25-27). Then, infectious diseases, cancer and mortality from all causes rise. This failing cell mediated immune response and proliferation of disease constitute another hallmark of aging.  High AA is necessary to prevent or reverse involution of the parenchymal mass of the thymus without which global synchrony of cell mediated immunity is lost; ie, cytotoxic T-cells against neo-antigens cannot be made, etc (28).

    **In Conclusion.** Scurvy is real whether overt or subclinical (mainly manifest by absence of AA in urine).  AA in the urine normally occurs in both synthesizers and non-synthesizers and it's absence is a first test for the existence of scurvy.

**References.**